\documentclass[iop]{emulateapj}

\shorttitle{Origin of Turbulence in Molecular Clouds}
\shortauthors{Tachihara et al.}

\begin{document}

\title{Toward Understanding the Origin of Turbulence in Molecular Clouds: \\
--- Small Scale Structures as Units of Dynamical Multi-Phase Interstellar Medium---}

\author{{Kengo Tachihara\altaffilmark{1, 2},}
{Kazuya Saigo\altaffilmark{2},}
{Aya E.~Higuchi\altaffilmark{1, 2},}
{Tsuyohshi Inoue\altaffilmark{3},}
{Shu-ichiro Inutsuka\altaffilmark{4},}
{Moritz Hackstein\altaffilmark{5},}
{Martin Haas\altaffilmark{5},}
{Markus Mugrauer\altaffilmark{6}}
}

\email{k.tachihara@nao.ac.jp}

\altaffiltext{1}{Joint ALMA Observatory, 
Alonso de C{\'o}rdova 3107, Vitacura, Santiago, Chile}
\altaffiltext{2}{National Astronomical Observatory of Japan, 
2-21-1, Osawa, Miaka, Tokyo, 181-8588, Japan}
\altaffiltext{3}{Department of Physics and Mathematics, Aoyama Gakuin University, 
Fuchinobe, Sagamihara 229-8558, Japan}
\altaffiltext{4}{Department of Astrophysics, Nagoya University, 
Chikusa-ku, Nagoya, 464-8602, Japan}
\altaffiltext{5}{Astronomisches Institut, Ruhr-Universit\"at Bochum, 
Universit\"atsstr.~150, D-44801, Bochum, Germany}
\altaffiltext{6}{Astrophysikalisches Institut und Universit\"ats-Sternwarte Jena, 
Schillerg\"a{\ss}chen 2-3, D-07745, Jena, Germany}

\accepted{May 19, 2012}

\begin{abstract}
In order to investigate the origin of the interstellar turbulence, detailed observations in the CO $J=$ 1--0 and 3--2 lines have been carried out in an interacting region of a molecular cloud with an \ion{H}{2} region.  As a result, several 1,000 to 10,000 AU scale cloudlets with small velocity dispersion are detected, whose systemic velocities have a relatively large scatter of a few km s$^{-1}$.  It is suggested that the cloud is composed of small-scale dense and cold structures and their overlapping effect makes it appear to be a turbulent entity as a whole.  This picture strongly supports the two-phase model of turbulent medium driven by thermal instability proposed previously.  On the surface of the present cloud, the turbulence is likely to be driven by thermal instability following ionization shock compression and UV irradiation.  Those small scale structures with line width of $\sim 0.6$ km s$^{-1}$ have a relatively high CO line ratio of $J=$3--2 to 1--0, $1 \la R_{3-2/1-0} \la 2$.  The large velocity gradient analysis implies that the 0.6 km s$^{-1}$ width component cloudlets have an average density of $10^{3-4}$ cm$^{-3}$, which is relatively high at cloud edges, but their masses are only $\la 0.05~M_{\sun}$.  

\end{abstract}


\section{Introduction}
Understanding the physical mechanism of contraction and condensation of molecular 
cloud cores toward star formation is a long standing issue.  
To date, two major paradigms have been 
proposed.  First is that cores are in a magnetically subcritical state, and as the 
magnetic field dissipates by ambipolar diffusion they collapse quasi-statically 
\citep[magnetic model;][]{shu87}.  The other model is that the interstellar medium (ISM) 
is dominated by supersonic turbulence and the collision of the turbulent flow forms 
temporary density enhancements, some of which undergo gravitational collapse 
\citep[turbulent model;][]{maclow04}.  The magnetic model is favored by some theoretical 
studies because the magnetic field can be modeled to have arbitrary strength and 
configurations (aligned or disordered).  It is, however, very difficult to be investigated 
observationally, being a 3-dimensional vector field.  On the other hand, turbulence 
is ubiquitous in interstellar space.  Almost all molecular line emission observed in 
molecular clouds have larger line widths than those expected from their thermal sound 
speed.  A difficulty of the turbulent model is its unknown nature, namely the origin of the 
turbulence.  The interstellar turbulence must have a much shorter dissipation timescale 
than lifetime of dense cores because its supersonic flow forms shocked regions and the 
energy is rapidly lost by radiation.  The turbulent model requires a driving mechanism 
to keep itself over a long timescale.  

These two models are supported by some observational evidence.  Low density diffuse 
clouds in general have filamentary structures, where aligned magnetic field lines (parallel 
or perpendicular to the long axis) are often observed \citep[e.g.,][]{myers91}.  These 
indicate that the ISM is frozen onto the magnetic field lines rather than bound by self-gravity.  
Molecular cloud cores are, on the other hand, reported to be dominated by turbulence, 
and those forming stars tend to have relatively small line widths \citep{tachihara02}.  
This suggests that as interstellar turbulence dissipates, cores tend to get gravitationally 
bound, condensed, and form stars.  It is also shown that the amount of turbulence is 
different among regions possibly due to difference in the environment; for example, cores 
in Taurus have much smaller line widths than those in Ophiuchus North and Lupus where 
star formation is not active and the timescale of core evolution is supposed to be longer 
\citep[e.g.,][]{tachihara00a,tachihara02}.  
Hence, turbulence is suggested to be one of the key elements controlling star formation
activity.  For these reasons, investigating the origin and driving mechanism of turbulence 
is vitally important for determining the timescale and necessary conditions of star formation.  
A clue may be in the environment, namely, many OB stars are distributed in the Ophiuchus 
North and Lupus regions where cores have relatively large line width.  A strong UV 
radiation field is known to affect kinematics and morphology of molecular clouds 
\citep[e.g.,][]{lefloch94}.

A promising theoretical model for the origin of interstellar turbulence was proposed by 
\citet{koyama00,koyama02} and following studies \citep{yamada07, inoue08, inoue09, 
hennebelle07a, hennebelle07b}.  Their model (hereafter the two-phase medium model) 
suggests that shock compressed layers of the interstellar 
diffuse gas undergo thermal instability and fragment into small scale ($\leq 1000$ 
AU) structures composed of denser cold neutral medium (CNM).  They are 
supposed to coalesce to form larger structures ($\sim 10000$ AU).  These cloudlets 
are embedded in warm neutral medium (WNM), and have random motions \citep[see 
Fig.~1 of][]{koyama02}.  The velocity is supersonic with respect to the sound speed 
of the CNM ($T \sim 50$ K), but subsonic with respect to that of WNM ($T \sim 8000$ K) 
\citep{field69,koyama00}.  Because of the two-phase medium, turbulence is sustained 
for relatively long timescales.  

In order to corroborate the idea of the two-phase medium model, we intend to detect small 
scale structures of the molecular cloud and investigate how interstellar turbulence is driven 
from the clouds' morphological and kinematical structures and their physical properties.  
We make detailed observations of CO $J=$1--0 and 3--2 at the surface of the molecular 
cloud \object{LDN 204} facing an \ion{H}{2} region where ionization shock compression 
is expected to be taking place. 

\section{Target region LDN 204}

\object{LDN 204} is a filamentary cloud complex facing the \object{Sh 2-27} \ion{H}{2} 
region ($r \sim 5$ pc), that is excited by the nearest O star, \object{$\zeta$ Oph} 
(SpT = O9.5V, $d = 140$ pc).  This cloud complex was entirely surveyed by the 
NANTEN telescope in $^{12}$CO  (HPBW = 2\farcm7), and it was found that the 
cloud is accelerated by illuminating UV radiation \citep{tachihara00b, liszt09}.  
The cloud complex horbours embedded dense cores discovered in C$^{18}$O, while star 
formation is not active and no associated young stars have been found \citep{tachihara00a}.  

The line width of C$^{18}$O $J=$1--0 observed by the NANTEN telescope in this 
region is typically as large as $\sim 0.74$ km s$^{-1}$, significantly larger than that 
in Taurus \citep[0.49 km s$^{-1}$ in average;][]{onishi96}.  
The strong UV radiation is suggested to have compressed the molecular cloud leading 
it to form dense cores, while the streaming motions of the low 
density gas imply considerable kinetic energy input to the ISM \citep{tachihara00b}.  
Noteworthy is that \object{$\zeta$ Oph} is a run-away star traveling through the region 
ejected from the center of the \object{Upper Sco} subgroup of the \object{Sco OB association}.  
Therefore the entire cloud complex is under the influence of the UV radiation from 
\object{$\zeta$ Oph} within a few $\times 10^5$ years.  

Another interesting aspect is that the filamentary cloud is penetrated by magnetic 
field lines perpendicular to the long axis of the filament \citep{mccutcheon86}.  
The cloud complex is suggested to be controlled by magnetic fields, while no further 
investigation toward the dense region has been reported.  

Because of the conditions mentioned above, in particular its proximity and the strong 
UV radiation, we chose the interface region of \object{LDN 204} and \object{Sh 2-27} 
as the best target for the present study.

\section{Observations}

\subsection{45 m telescope observations}

We carried out CO $J=$1--0 observations at 115 GHz with the NRO 45 m telescope 
(HPBW = 15\arcsec, corresponding to 2000 AU).  The interface region of the cloud 
surface of $11\arcmin \times 22\arcmin$ area centered at (R.A., Dec.)$_{J2000}$ = 
(16:46:45.0, $-12$:21:30.0) is surveyed with the BEARS multi-beam receiver 
employing the On-The-Fly (OTF) mapping mode.  
We repeated the vertical and horizontal scanning many times and summed the data 
afterward in order to avoid scanning effects.  
We also employed scaling for correcting the beam efficiency by multiplying the 
beam-specific factors provided by the Nobeyama observatory.  
The telescope pointing was checked every 2 hours by obtaining 5-point 
observations toward nearby SiO maser sources.  

Throughout the observing runs, the wind speed has been below 15 m s$^{-1}$.  
About one fifth of the observations were taken under slightly windy conditions as wind 
speed was $\ge 10$ m s$^{-1}$.  We thus divide the data into two to see 
how the spatial resolution gets worse under the windy conditions.  No significant 
difference between windy and calm conditions can be seen, and eliminating 
the data with windy conditions worsens the S/N ratio of the final image.  We therefore 
decided to use all the data.  

The data taken with the OTF mapping is reduced on the NRO data reduction software 
called ``NOSTAR" \citep{sawada08}.  The baseline is subtracted by fitting to a linear 
function.  The data is re-gridded into a 7\farcs5 grid spacing by convolving with 
the first order Bessel function $\times$ the Gaussian function, resulting in the final data 
cube with an effective spatial resolution of 19\farcs5.  
For each spectrum of the 7\farcs5 grid cell, a final signal sensitivity of $T_{\rm rms} 
= 0.37$ K with a velocity resolution of 0.1 km s$^{-1}$ is achieved.  

\subsection{ASTE observations}

Toward three selected regions where peculiar small scale structures are detected, we made 
follow-up CO $J=$3--2 observations with the Atacama Submillimeter Telescope Experiment 
(ASTE).  It has a heterodyne receiver working at 345 GHz with typical noise temperature of 
$T_{\rm sys} \sim 400$~K (double-sideband) at the zenith.  
It is equipped with the 1024-channel digital backend that has a frequency resolution of 125 kHz, 
which corresponds to the velocity resolution of 0.11~km s$^{-1}$.  
Its 10 m dish enables us to have comparable spatial resolution to the 45 m telescope with the 
HPBW of 22\arcsec.  
The two target regions are centered at (16:46:45.0, $-12$:22:50) and (16:46:36.5, $-12$:28:30) 
with sizes of $4\arcmin \times 4\arcmin$ and $3\arcmin \times 3\arcmin$, respectively, 
and are mapped with the OTF mode.  
The pointing error was corrected every 2 hours measuring the offset from a strong point-like 
source in CO.  The calibration for the beam efficiency is done by observing spectra of 
M17SW (R.A., Dec.)$_{J2000}$ = (18:20:23.1, $-16$:11:43) assuming that the peak radiation 
temperature ($T_{R}^{*}$) is 69.6~K \citep{wang94}.  
The data are reduced with NOSTAR, similar to the 45 m data, resulting in the final data 
cube with the effective spatial resolution of 24\farcs2.
The rms noise temperature ($T_{\rm rms}$) of the spectra of each 7\farcs5 grid cell 
are 0.13 K after calibration.  

\section{Results}

\subsection{Spatial and velocity structures in $J=$1--0}

\begin{figure*}
\epsscale{0.8}
\plotone{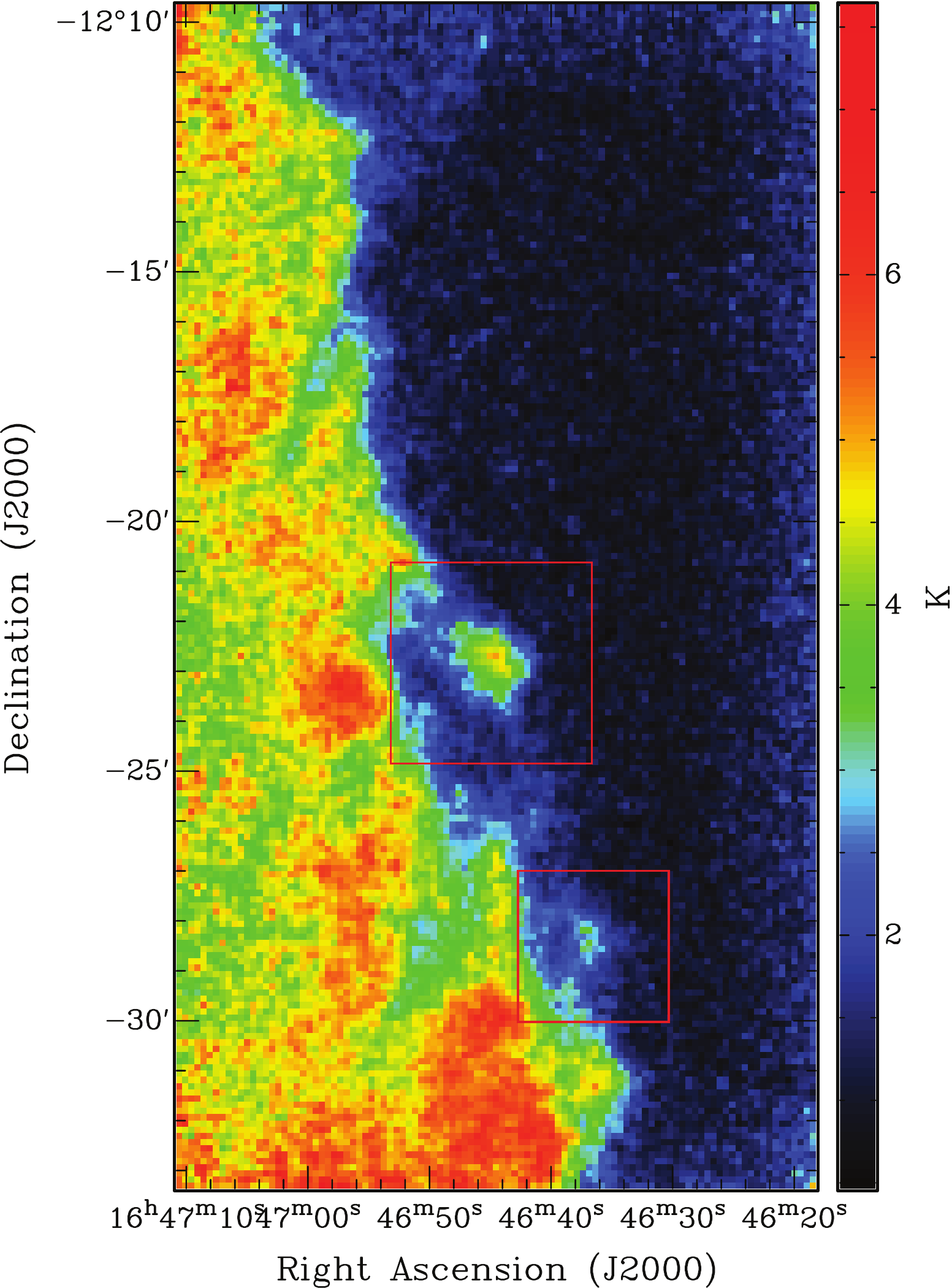}
\caption{Pseudo color image of peak $T_{\rm MB}$ map in CO $J=$ 1--0 observed with the NRO 
45 m telescope.  The two red squares denote the regions where CO $J=$3--2 data are 
taken with ASTE.  
}
\label{peakT}
\end{figure*}

\begin{figure*}
\epsscale{0.7}
\plotone{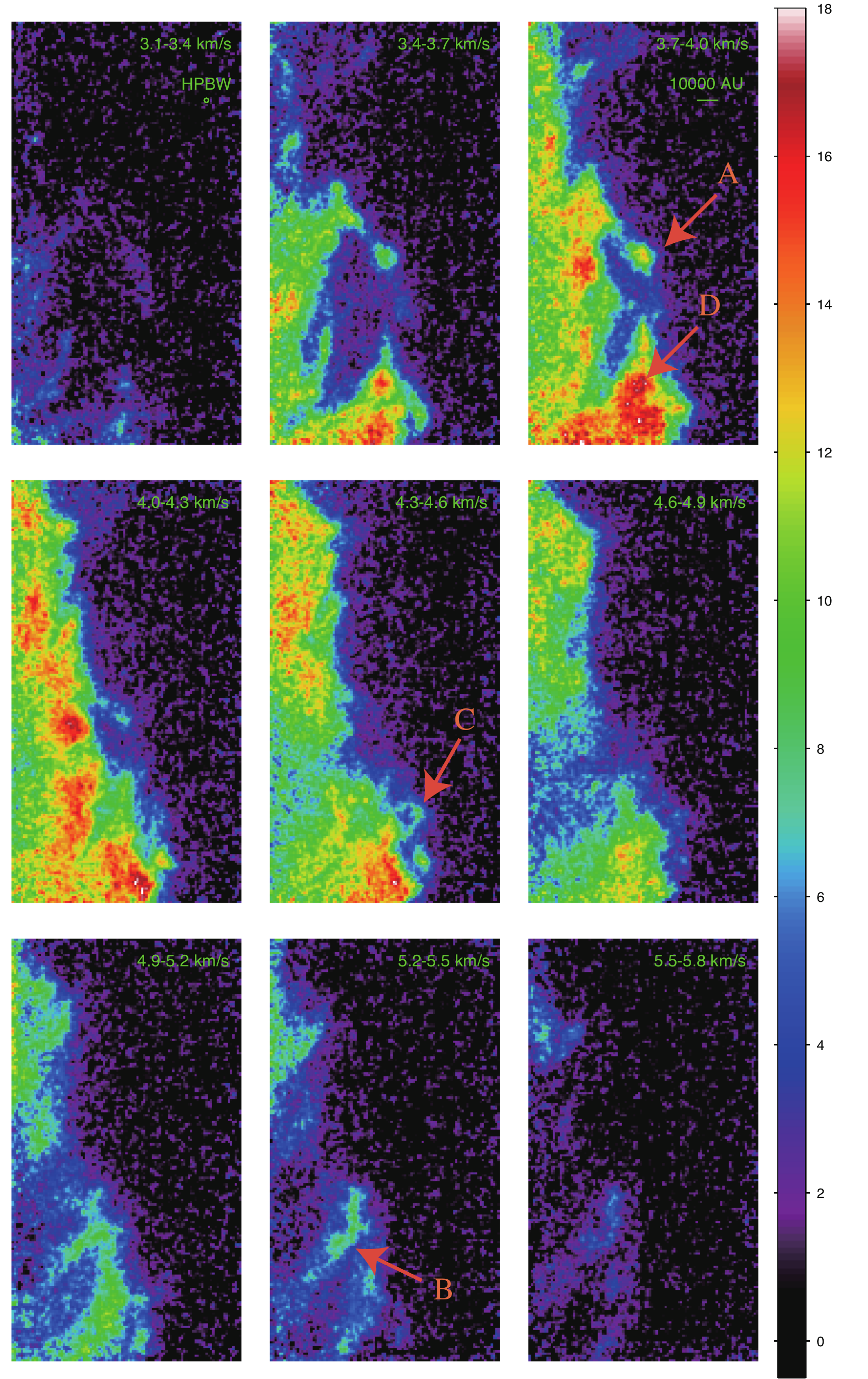}
\caption{Channel map of the CO $J=$ 1--0 emission for the same region as Fig.~\ref{peakT}.  
Each map shows CO intensity 
integrated over 0.3 km s$^{-1}$ velocity width as shown at the top right of each 
map.  Three prominent structures (the clump, pillar and arc) are designated by A, B, and C, 
respectively, in addition to a middle part of the cloud by D.  The data used to create 
this figure are available in the online journal.
}
\label{channelmap}
\end{figure*}

Fig.\ \ref{peakT} is the peak $T_{\rm MB}$ map of the CO $J=$1--0 line in the cloud boundary 
region of \object{LDN 204}.  The west side of the cloud is facing the \ion{H}{2} region 
\object{Sh 2-27} and the cloud surface is illuminated by UV flux from \object{$\zeta$ Oph} 
located in the direction of north west.  The edge of the cloud does not appear to be a 
smooth surface, but has a complex morphology, for example, there is a prominent clumpy 
structure protruding to the west in the middle of the observed area.  
The velocity field is also remarkably complex as seen in the channel maps of 
Fig.\ \ref{channelmap}.  

Particularly a few prominent features can be recognized such as 
clumpy, pillar-like, and arc-like features as denoted by A, B, and C in Fig.\ \ref{channelmap}, 
respectively.  The clumpy structure is especially noticeable in the peak $T_{\rm MB}$ 
map as slightly detached from the main cloud.  The clump (A) has a radius of $\sim 45\arcsec$, 
corresponding to $r \sim 6300$ AU.  Using the empirical relation of 
$X \equiv N({\rm H_2}) / W({\rm CO}) = 1.56 \times 10^{20}$ cm$^{-2}$ (K km s$^{-1}$)$^{-1}$ 
\citep{hunter97}, 
the total mass of the clump is estimated to be 0.05 $M_\sun$, where $N({\rm H_2})$ is 
the column density of molecular hydrogen and $W({\rm CO})$ is the integrated intensity 
of the CO $J=$ 1--0 line.  Similar morphological structures have 
been discovered at interaction regions of the cloud surfaces with \ion{H}{2} regions like 
\object{M16} \citep[e.g.,][]{hester96}.  They are referred to as ``Evaporated Gaseous 
Globules" (EGGs).  The size of the clump is, however, much larger than those of EGGs 
that range from 150 AU to 1800 AU.  
The pillar-like feature (B) is visible in the channels around $V_{\rm LSR} \approx 5.4$ 
km s$^{-1}$ and overlapping with other velocity components.  
It is roughly pointing to the direction of the UV radiation source of \object{$\zeta$ Oph}.  
The arc-like feature (C) is protruded from the main cloud by $\sim 7000$ AU, and if it is 
assumed to be a bent cylinder, its radius is $r \sim 3000$ AU.  

\subsection{Velocity dispersions of the small scale structures}
\label{vdisp}

\begin{figure}
\epsscale{1}
\plotone{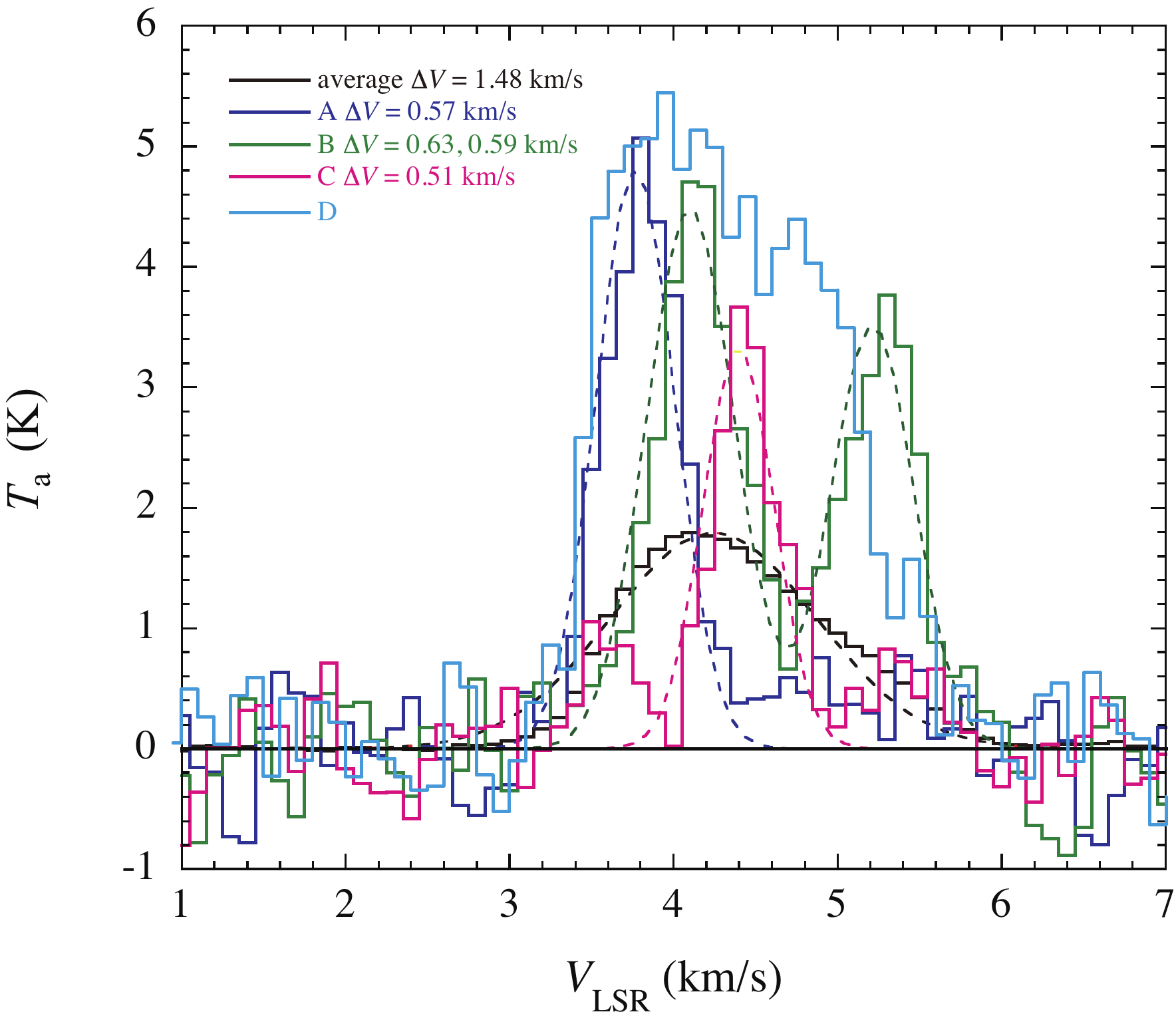}
\caption{Spectra of the CO $J=$ 1--0 line at the 3 peak positions of the clumpy (A in blue), 
pillar-like (B in green), arc-like (C in magenta) structures, and the middle of the cloud (D 
in cyan) as shown in Fig.~\ref{channelmap}.  
The black curve illustrates the average CO $J=$ 1--0 spectrum of the entire observed region.  
The dotted curves are the gaussian profiles fitted to those spectra, whose line width (FWHM) 
are indicated in the top-left of the figure.  }
\label{spectra10}
\end{figure}

The CO $J=$1--0 line profiles at the peaks of the 3 prominent features and a typical line 
profile in the middle of the cloud are investigated.  
Fig.\ref{spectra10} illustrates that the lines at the 3 positions are fitted to single or double 
gaussian profiles and the central velocities and velocity dispersions are estimated.  
For positions A and C, the line emissions are fitted to single gaussians whose Full Width 
Half Maximum (FWHM) line widths are 0.57 km s$^{-1}$ and 0.51 km s$^{-1}$, respectively.  
At position B, the line has clear double components, which have FWHM of 0.63 km s$^{-1}$ and 
0.59 km s$^{-1}$.  The central velocities are clearly different ranging from 3.8 km s$^{-1}$ 
to 5.3 km s$^{-1}$.  On the other hand, in the middle of the cloud, the spectral line shapes 
are not well fitted to the gaussian profile, but more irregular.  They seem to be saturated 
at typical peak antenna temperature of $\sim 6$ K, and consist of multiple velocity 
components overlapping on the line of sight forming large velocity dispersion.  
The composite average spectrum over the entire observed area is illustrated by the 
black line in Fig.~\ref{spectra10}.  The gaussian fitting to it gives a much larger line width 
of $\sim $ 1.5 km s$^{-1}$ than those of the small-scale structures (hereafter 0.6 km 
s$^{-1}$ component).

The observed line width is composed of thermal and non-thermal components expressed 
as $\Delta V_{\rm obs}^2 = \Delta V_{\rm th}^2 + \Delta V_{\rm NT}^2$, where the thermal 
component of the velocity dispersion at temperature $T$  is expressed as 
$\Delta V_{\rm th} = \sqrt{8 \ln 2 \cdot k T /m}$ where $k$ is the Boltzman's constant 
and $m$ is the mass of the molecule.  For the cases of $T=20, 60, 150$ K, $\Delta V_{\rm th}$ 
of the $^{12}$CO molecule is calculated to be 0.2, 0.35, 0.55 km s$^{-1}$, respectively.  
Because we do not expect temperatures as high as 1000~K for molecular clouds, typical 
line widths of $\sim 1.5$ km s$^{-1}$ have been interpreted as turbulent rather than 
thermal motion. The velocity dispersion of $\sim 0.6$ km s$^{-1}$ implies, on the 
contrary, that the cloud small structures are nearly thermalized with temperature of 
$\sim 150$ K, or they have relatively small non-thermal components.  

These results imply that the molecular cloud is not made of a uniform material, but 
composed of small-scale cloudlets whose internal velocity dispersion is as small as 
the thermal motion of CNM, while the relative motion 
of the individual cloudlets is much larger.  Because many small-scale cloudlets are 
overlapping on the same line of sight in the main cloud, the spectral lines obtained 
in the middle of the cloud are widened and exhibit larger line width mimicking supersonic 
motion of the entire material in the cloud.  The relative motion of the cloudlets 
is in fact in the order of 1-2 km s$^{-1}$, which is supersonic with respect to the 
temperature of CNM ($T \sim 50$ K), but subsonic with respect to WNM ($T \sim 8000$ K).  
This interpretation is consistent with the two-phase medium turbulent model of \citet{koyama02}.  

\subsection{CO $J=$3--2 results}

The data of ASTE CO $J=$3--2 and 45 m $J=$1--0 are compared in detail for the purpose 
of investigating the gas density and temperature of the small-scale cloud structures.  
In order to make a point-to-point comparison, the 45 m data are spatially convolved with 
a gaussian beam of 22\arcsec\ HPBW so that the 2 datasets have the same spatial resolution.  

\begin{figure*}
\epsscale{0.5}
\plotone{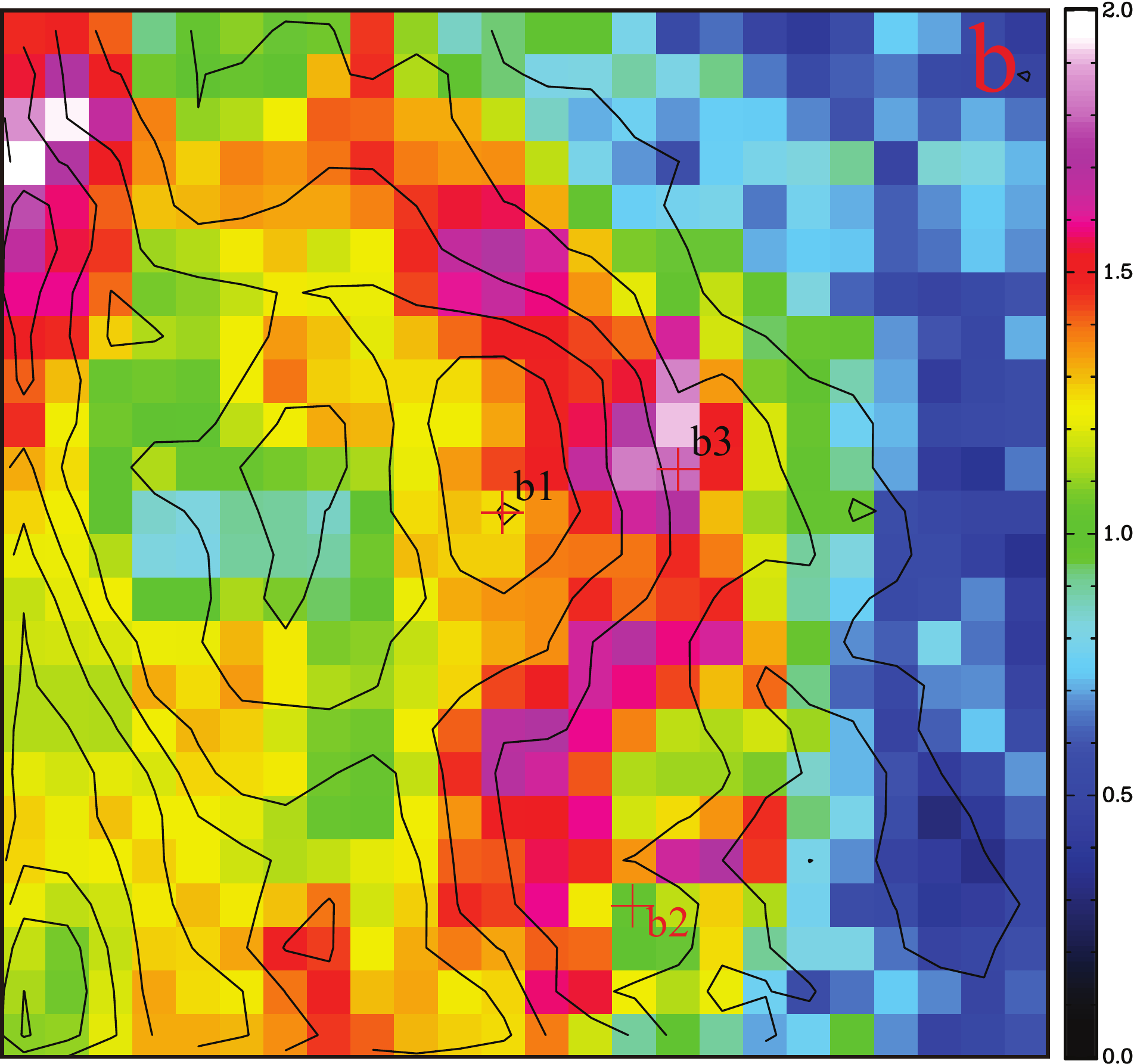}
\plotone{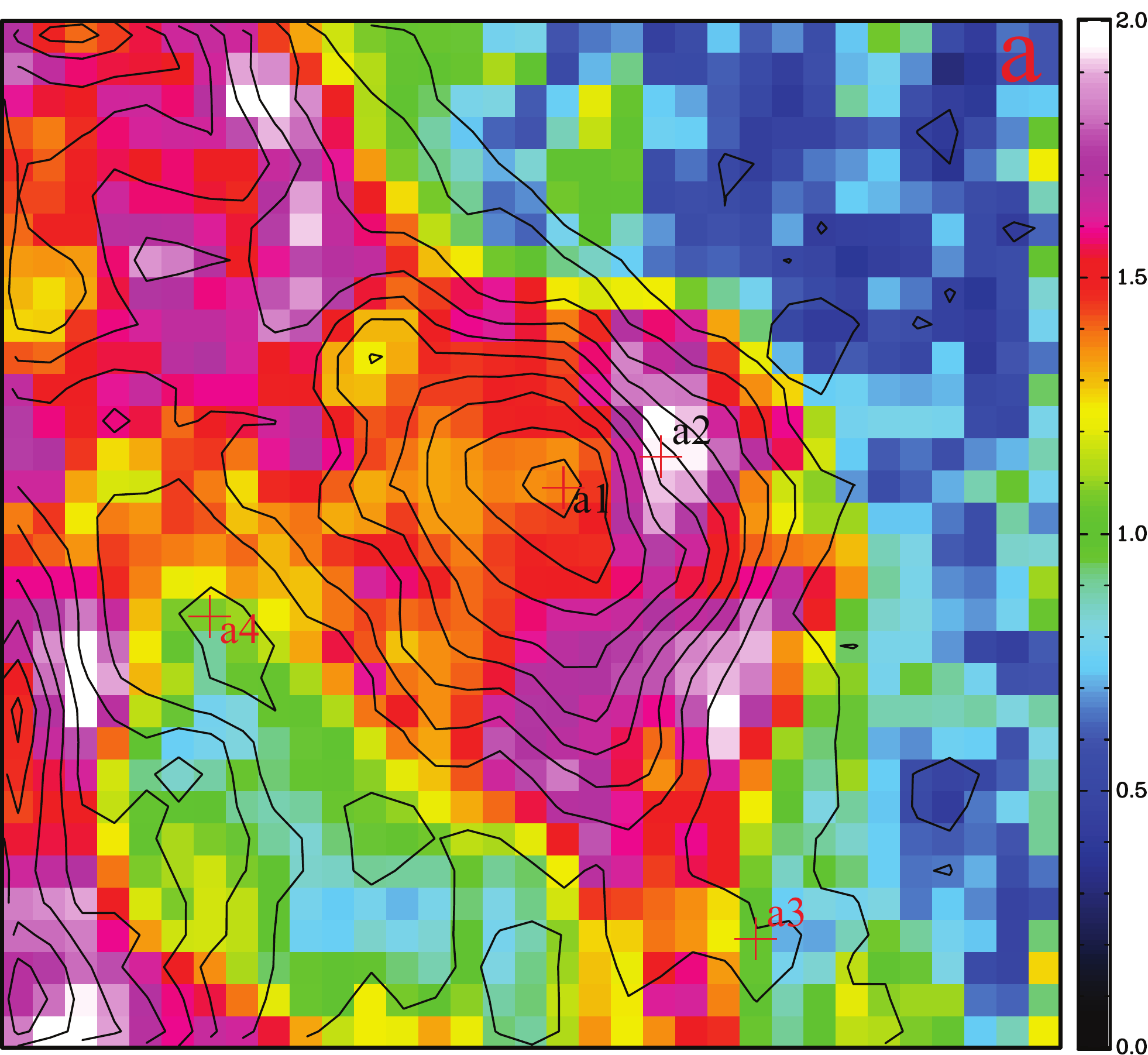}
\caption{Pseudo color image of the peak $T_{\rm MB}$ ratio of $R_{3-2/1-0}$ around the clump (a), 
the arc (b), respectively.  Overlaid are contours of 
peak $T_a\ J=$ 1--0 from 1.0 K ($= 2.7\sigma$) with 0.4 K steps.  
The coordinates of the center positions are (16:46:45.0, -12:22:50) and 
(16:46:36.5, -12:28:30), respectively, and the sizes of the maps are 
$4\arcmin \times 4\arcmin$ and $3\arcmin \times 3\arcmin$, respectively, as denoted 
by the red boxes in Fig.~\ref{peakT}.  
}
\label{ratio}
\end{figure*}

\begin{figure*}
\epsscale{0.55}
\plotone{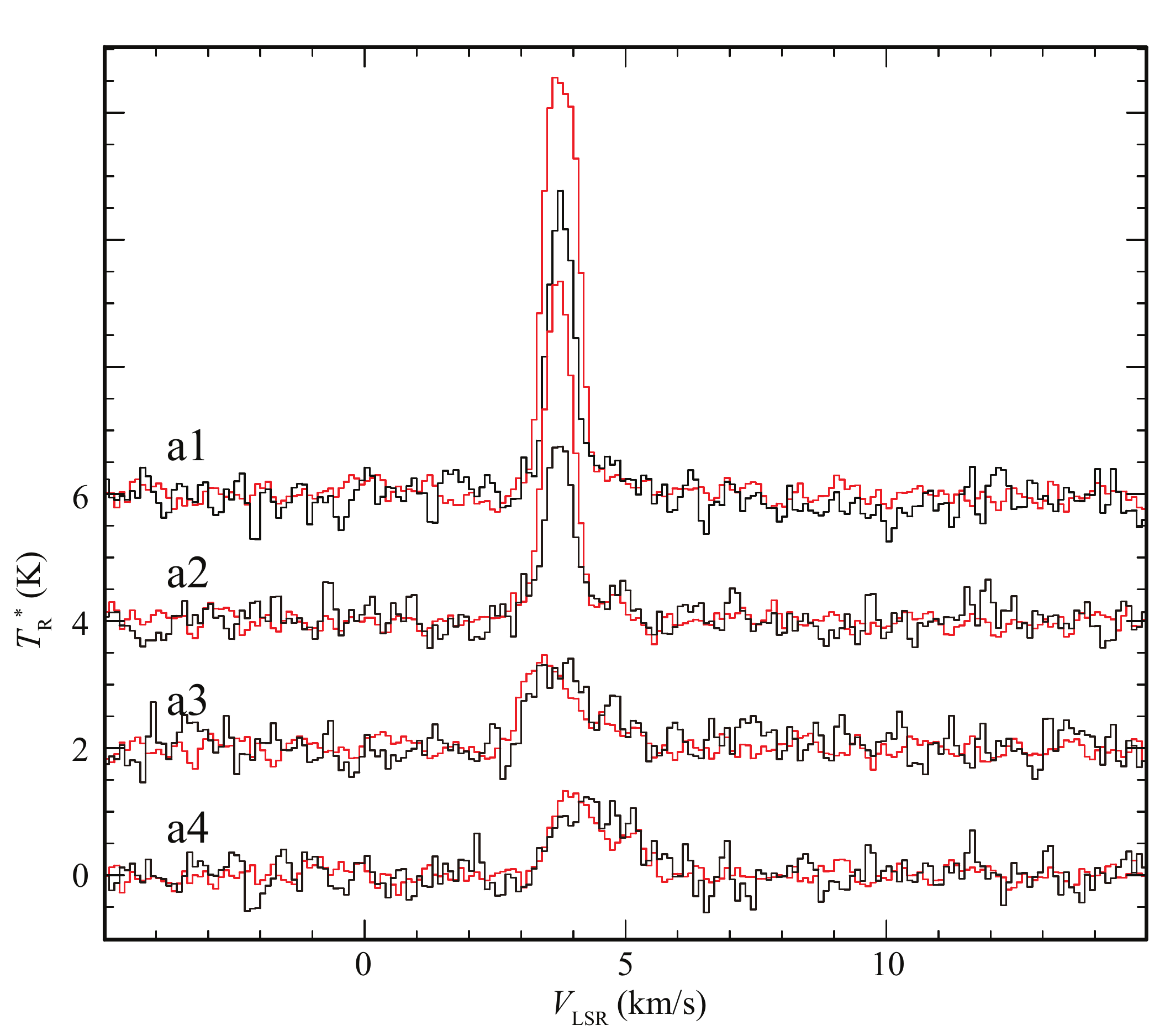}
\plotone{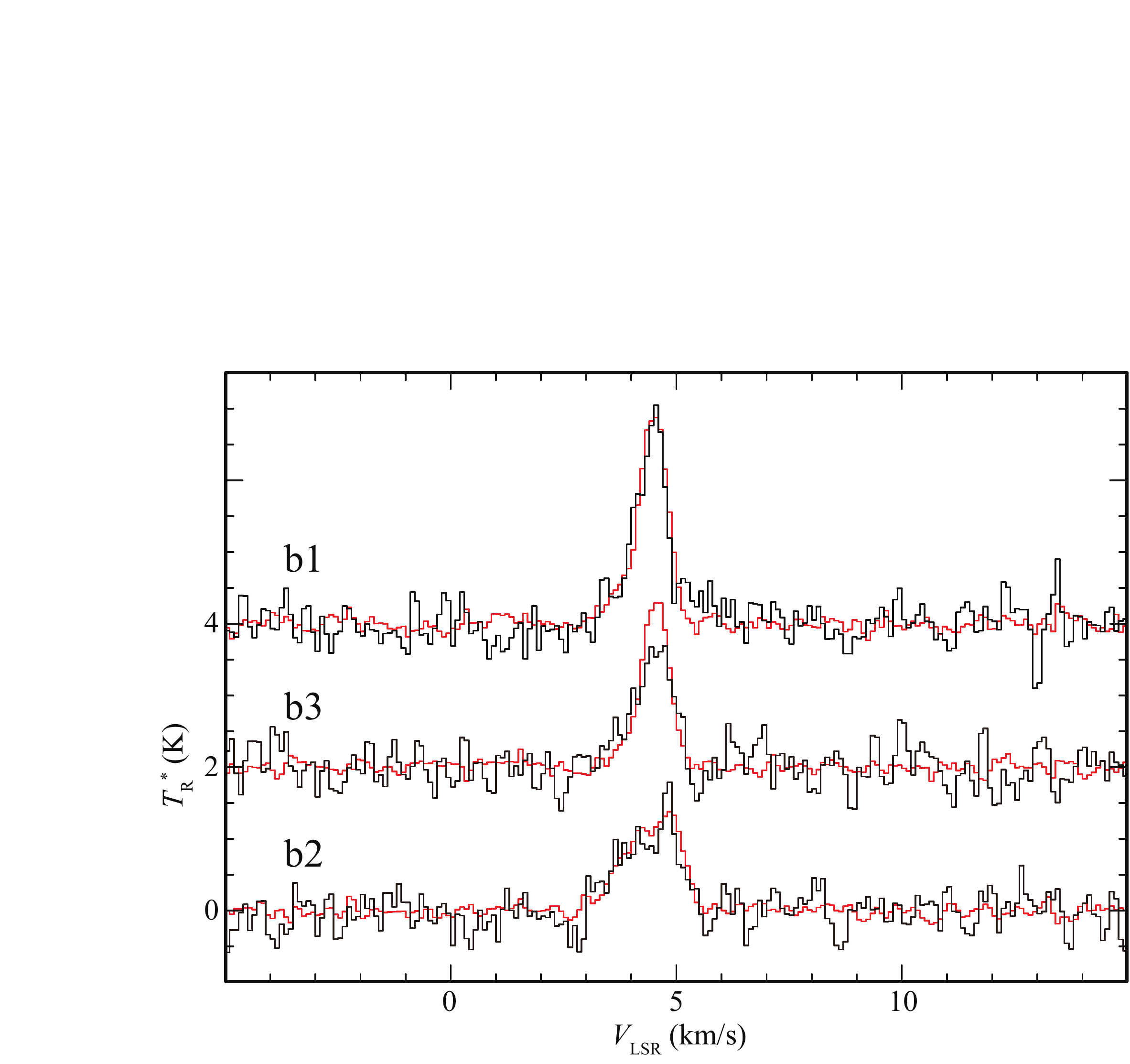}
\caption{Spectra of CO $J=$ 1--0 (black) and 3--2 (red) at the positions marked in 
Fig.~\ref{ratio}.  The CO $J=$ 1--0 data are convolved with a gaussian beam of 
22\arcsec\ HPBW so that the 2 data cubes have the same spatial resolution.  
The intensity and the spectral shapes of the emission lines of $\Delta V \sim 0.6$ 
km s$^{-1}$ are similar for both transitions except for a1 and a2.
The data used to create Figures \ref{ratio} and \ref{spectra32} are available in 
the online journal.
}
\label{spectra32}
\end{figure*}

Fig.~\ref{ratio} illustrates the peak $T_{\rm MB}$ ratio of $R_{3-2/1-0}$ for the 2 regions 
where (a) the clumpy and (b) the arc-like features are detected.  In region a, the clumpy 
feature entirely has $R_{3-2/1-0} \geq 
1$ and at the peak position a1, $R_{3-2/1-0} \approx 1.4$.  The west side of the clump 
facing the \ion{H}{2} region has even higher ratio of $R_{3-2/1-0} \approx 2$ at position 
a2.  The west side of the clump has a steep intensity gradient and is illuminated by strong 
UV radiation, so the gas temperature is expected to be higher than the eastern side.  
Both of the $J=$1--0 and 3--2 lines at positions a1 and a2 have line widths of $\sim 0.6$ 
km s$^{-1}$, while in the periphery at a3 and a4, their line widths are significantly 
larger than $\sim 1.5$ km s$^{-1}$ (Fig.~\ref{spectra32}a).  
In region b, the arc-like feature shows a similar trend of $R_{3-2/1-0}$ having higher 
values in the west side.  At the peak position b1, $R_{3-2/1-0}$ is close to unity.  
The $J=$3--2 lines' intensity and shapes resemble the $J=$1--0 lines in general as 
seen at positions b1, b2, and b3.  

\section{Discussion}

\subsection{Physical properties of the small scale molecular cloud}

In the observed region, there are small scale structures  detected near the main cloud boundary, i.e., the 0.6 km s$^{-1}$ width component whose  typical size is $6000~{\rm AU} \la l \la 12600~{\rm AU}$.   According to the theory of \citet{koyama02}, thermal instability of shock-compressed  WNM ($T \sim 8000$ K) forms thermalized cloudlets of a few 100 AU made of CNM  ($T \sim 50$ K) that are embedded in the WNM.  They have random motion with a velocity dispersion of a few km s$^{-1}$ and coalesce with each other to  form larger structures up to several $\times 1000$ AU.   Our results are consistent with this scenario from the morphological and kinematical  point of view.  In order to further corroborate the theory, estimation of gas density and temperature are attempted in the following.  

The 0.6 km s$^{-1}$ width component is well traced by both of the CO $J=$1--0 and 3--2  lines.   The gas density and temperature can be constrained from the intensities of multiple lines by the large velocity gradient (LVG) analysis, which requires density, temperature, column density, abundance,  and velocity gradient.  For the calculation, we adopt flowing assumptions setting the density  ($n({\rm H_2})$) and the kinetic temperature ($T_{\rm kin}$) as free parameters.   1) H$_2$ density ($n({\rm H_2})$), CO abundance ratio to H$_2$ ($Z$), and column density ($N({\rm CO})$) have the relation of $N({\rm CO}) = Zn({\rm H_2})l$  where $l$ is the path length of the cloud along the line of sight.   2) The path lengths ($l$) are the same as the apparent diameters of the clouds ($2 \times r$), i.e.,  spherical or axial symmetry.   3) The velocity gradients are the same as the CO $J=$1--0 line widths divided by the path lengths  ($\Delta V / l$).  4) The CO abundance is constant as $Z \sim 3 \times 10^{-5}$ for the entire clouds.   The last assumption is, however, unlikely because the present region is under strong UV radiation and a considerable amount of CO molecules is dissociated to be \ion{C}{1}, particularly at the cloud surface.  We therefore tested the calculation altering the CO abundance with 3 different $Z$ values of $5 \times 10^{-5},\ 1 \times 10^{-5}$, and  $5 \times 10^{-6}$.   As a result, it is found that the response of $R_{3-2/1-0}$ to the same $n({\rm H_2})$ and $T_{\rm kin}$ changes with $Z$ by a factor of 3 or less, while the velocity integrated CO intensity, $W({\rm CO})$, changes up to an order of magnitude.  

\begin{figure*}
\epsscale{0.55}
\plotone{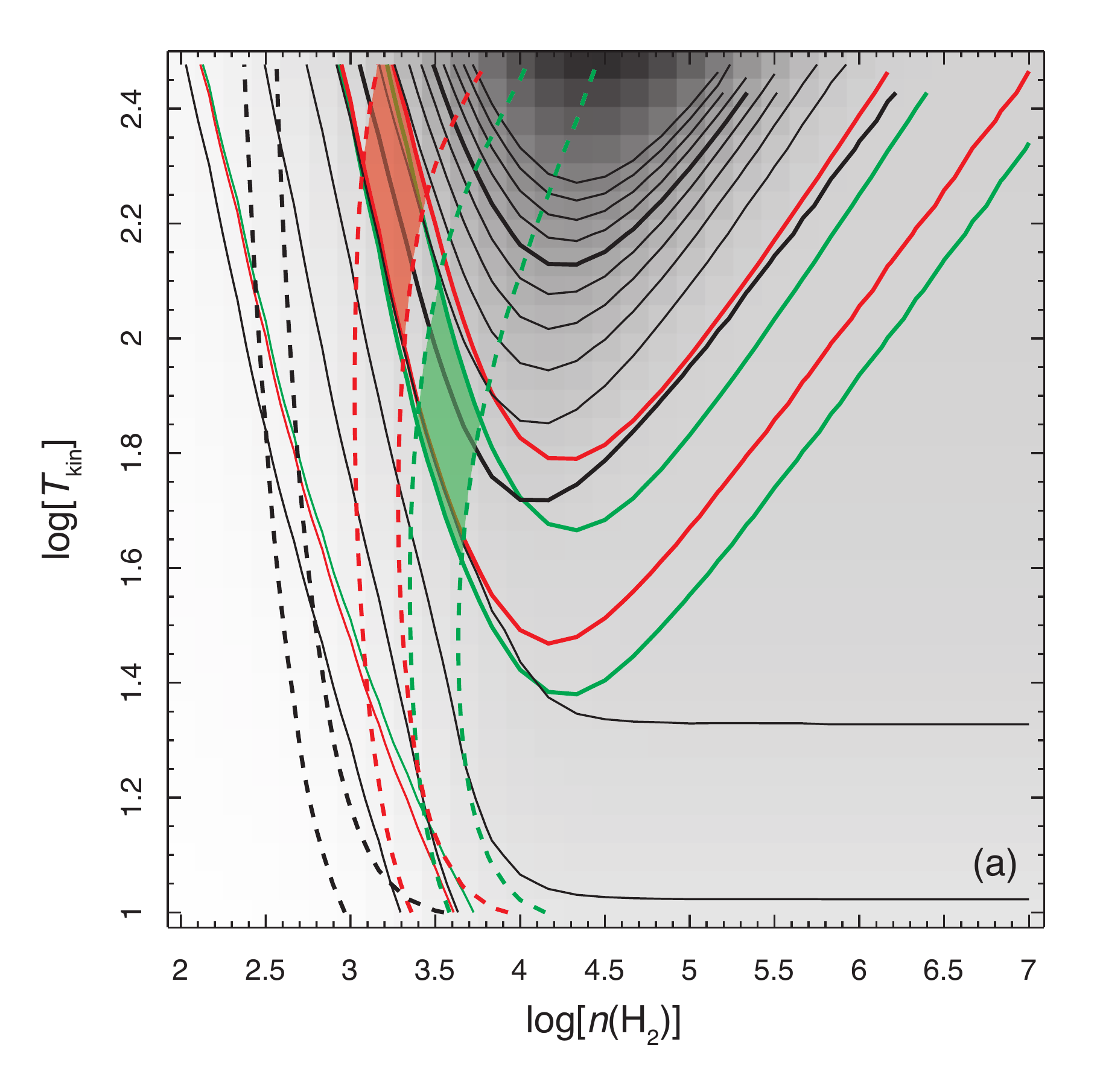}
\plotone{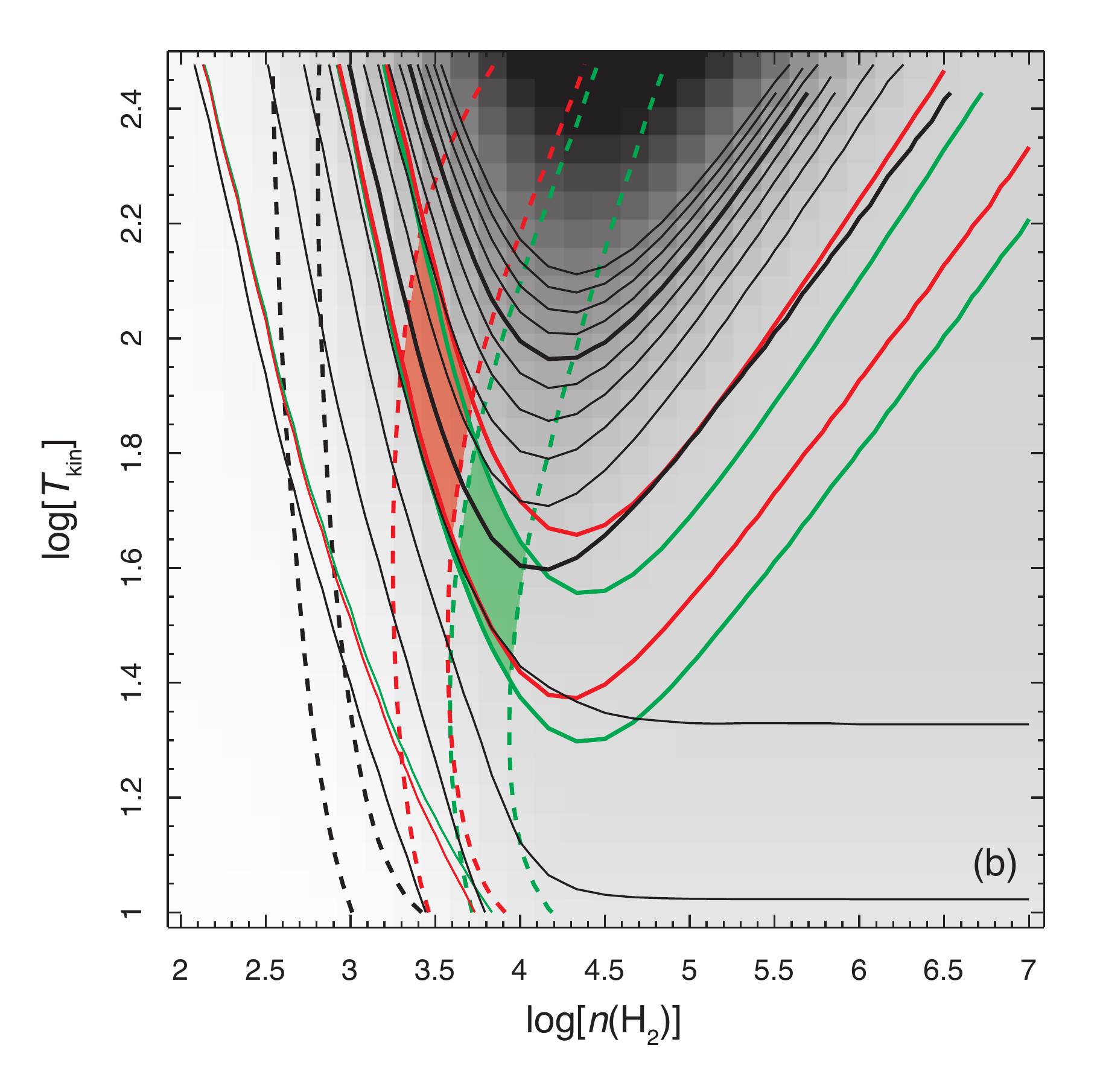}
\caption{Intensity ratio diagrams of CO $R_{3-2/1-0}$ as a function of molecular 
hydrogen density ($n({\rm H_2})$) and kinetic temperature ($T_{\rm kin}$) shown by 
the gray scale and contours assuming the CO abundance of $5 \times 10^{-5}$.  
Contours are from 0.2 to 2.8 with steps of 0.2, and the bold contours indicate the line 
ratio of unity and 2.  The red and green lines are contours of 0.2, 1.0 and 2.0 for the 
cases of lower CO abundance with $Z = 1 \times 10^{-5}$ and $5 \times 10^{-6}$, 
respectively.  Used parameters are (a) $l=12600$ AU and $\Delta V = 0.6$ km s$^{-1}$ 
and (b) $l=6000$ AU and $\Delta V = 0.6$ km s$^{-1}$.  
The calculated integrated intensity of CO $J=$1--0 under the given condition are 
shown by dotted contours of 2.5 and 4.0 K km s$^{-1}$ for (a) and 
2.0 and 3.5 K km s$^{-1}$ for (b), respectively. 
The shaded areas indicate the parameter ranges meeting the observed values 
(see text).
}
\label{lvg}
\end{figure*}

Figs.~\ref{lvg} illustrate the results of the LVG calculations showing $R_{3-2/1-0}$ as a function of $n({\rm H_2})$ and $T_{\rm kin}$ by gray scale and contours.  Fig.~\ref{lvg}a is for the case of $\Delta V = 0.6$ km s$^{-1}$ and $l=12600$ AU, corresponding to the clumpy structure, which demonstrates the line ratio is in the range of $1 < R_{3-2/1-0} < 2$ for the entire region, 
while the integrated intensity $W({\rm CO})$ in the clumpy structure is between 2.5 and 4.0 K km s$^{-1}$ for the $J=$1--0 line.  The parameter set of $n({\rm H_2})$ and $T_{\rm kin}$ that meets these observed conditions does not exist in this figure with $T_{\rm kin} < 300$ K for the case of $Z = 5 \times 10^{-5}$.  For the lower abundance cases, these conditions are satisfied with the parameter in the shaded areas of Fig.~\ref{lvg}a.  For the case of $Z = 1 \times 10^{-5}$, the density and kinetic temperatures are restricted to be $1000 \la n({\rm H_2}) \la 3000$ cm$^{-3}$ and $100 \la T_{\rm kin} \la 300$ K; and for $Z = 5 \times 10^{-6}$, they are $2500 \la n({\rm H_2}) \la 6000$ cm$^{-3}$ and $40 \la T_{\rm kin} \la 120$ K.  These are warmer and denser compared to typical dark clouds detected in $^{12}$CO.  From the size and thus constrained density, the mass of the clump is given as $0.008 \la M \la 0.05~M_\sun$, which is below the brown-dwarf mass limit.  Since these masses are well below the Jeans mass at their current densities, these cloudlets are not expected to collapse gravitationally.  Their further evolution may include mutual coalescence, condensation from surrounding medium, ``evaporation" (from cold gas to warm gas), and photo-dissociation.

Fig.~\ref{lvg}b is for $l = 6000$ AU and $\Delta V = 0.6$ km s$^{-1}$ as for the arc-like structure, illustrating the similar trend, but using the same $n({\rm H_2})$ and $T_{\rm kin}$, $R_{3-2/1-0}$ is calculated to be higher and $W({\rm CO})$ is lower than Fig.~\ref{lvg}a, yielding lower kinetic temperature for the same $R_{3-2/1-0}$ value and require higher density for the same $W({\rm CO})$.  Note that in the area on the left side of the turn-over point of the $R_{3-2/1-0}$ contours, both lines are optically thin.  
As seen in Fig.~\ref{ratio}b, the arc-like structure exhibits higher $R_{3-2/1-0}$ than unity for the entire structure, and the peak ratio is $\sim 2$ around the position b4.  
The integrated intensity of the $J=$1--0 line in this structure is in the range between 2.0 and 3.5 K km s$^{-1}$.  Hence, as for Fig.~\ref{lvg}a, there is no region meeting the observed constraints with $T_{\rm kin} < 300$ K and $Z = 5 \times 10^{-5}$, while we obtained a satisfying region for the lower abundance cases.  For the case of $Z=1 \times 10^{-5}$, the density and kinetic temperatures are constrained as $2000 \la n({\rm H_2}) \la 5000$ cm$^{-3}$ and $50 \la T_{\rm kin} \la 150$ K; and for the case of $Z = 5 \times 10^{-6}$, $4000 \la n({\rm H_2}) \la 11000$ cm$^{-3}$ and $25 \la T_{\rm kin} \la 60$ K.  These imply that the arc-like structure is cooler and denser than the clumpy one.  From the derived density range, the mass of the arc-like structure is also constrained to be very low-mass as $0.005 \la M \la 0.03~ M_{\sun}$.  

If we use the empirical relation of $N({\rm H_2}) = X \cdot W({\rm CO})$ with an assumption of $X = 1.56 \times 10^{20}$ cm$^{-2}$ (K km s$^{-1}$)$^{-1}$ \citep{hunter97}, the average density is thus derived only from $W({\rm CO})$ and $l$ as $\sim 3300$ cm$^{-3}$ and $\sim 6000$ cm$^{-3}$ for the clumpy and arc-like structures, respectively.  Under the circumstance with strong UV radiation, smaller $Z$ and thus larger $X$ are expected than above, yielding larger density, and therefore consistent density estimations are plausible.  Nevertheless, these estimates are uncertain under such an extreme condition.  

Although the mean kinetic temperature of molecular gas along the line of sight toward \object{$\zeta$ Oph} was estimated to be 54 K from H$_2$ absorption \citep{savage77, liszt09}, small scale structures may have large temperature variations.  The estimated kinetic temperature of the 0.6 km s$^{-1}$ component is relatively high compared to typical dark clouds, but they are still much colder than WNM.
\citet{tachihara00b} reported that the entire \object{LDN 204} cloud complex is a 20-pc long and 2-pc wide filament with an estimated mass of $1100~M_{\sun}$, yielding the average density of $\sim 250$ cm$^{-3}$.  The above calculations reveal the clumpy and arc-like structures to be 4 to 40 times denser than the cloud average.  

\subsection{Other observational evidence of small-scale cloud structures}

From observational points of view, some studied have been  published attempting to investigate small-scale structures of ISM.  \citet{sakamoto02} and \citet{sakamoto03} detected similar small-scale structures at the edge of high-latitude clouds (\object{MBM 54} and \object{MBM 55}) and \object{Heiles Cloud 2} in Taurus, respectively, although their observations were by one-dimensional strip scan.  The typical one-dimensional size is $\sim 10000$ AU, comparable to our 0.6 km s$^{-1}$ structures, while their FWHM line widths are $\simeq 1.8$ and $\simeq 2.8$ km s$^{-1}$ for \object{MBM 54} and \object{MBM 55}, and  $\simeq 2$ km s$^{-1}$ for \object{HCL2}, respectively, significantly larger than the present results.  Without strong UV radiation, these regions are supposed to have smaller pressure and remarkably low density of $n ({\rm H_2}) \leq 100$ cm$^{-3}$ is estimated, although no detailed estimation of the physical cloud parameter was performed.  These clouds with different velocity widths may have been caused by different initial conditions.  

\citet{falgarone09} investigated even smaller-scale structures using the IRAM-PdB interferometer in the Polaris Flare, and detected elongated structures of $\sim 300$ AU thickness.  They tend to be seen in the CO $J=$1--0 line wings with line width of 0.1-0.4 km s$^{-1}$.  Because some of the elongated structures form parallel pairs with different velocity, it was suggested that the small structures are thin layers of CO cloud with velocity shears driven by turbulent flow.  

At high galactic latitude, small-scale CO structures have been discovered by single-dish survey and by interferometry.  \citet{heithausen02} serendipitously discovered small-scale ($< 1\arcmin$) structures with $\Delta V \sim 0.8$ km s$^{-1}$ with the IRAM 30m telescope.  The clouds are resolved into smaller-scale ($\sim$ a few 10\arcsec) structures with $\Delta V \sim 0.4$ km s$^{-1}$ by the PdBI interferometric observation \citep{heithausen06}.  Assuming the distance of 100 pc, their sizes and densities are $\leq$ a few $\times 100$ AU and $\geq 20,000$ cm$^{-3}$, respectively.  By a follow-up large-scale survey, \citet{heithausen06} claimed that such small clouds are common in the ISM, but their total mass amounts to only $< 1\%$ of the whole gas mass.  The size and $\Delta V$ of the small-scale structures ranging from $\sim 100$ AU to $10^4$ AU including an ensemble cloud have a correlation of $\Delta V \propto r^{0.3}$, where $r$ is the effective radius of the cloud.  Our sample of the 0.3 km s$^{-1}$ and the 0.6 km s$^{-1}$ width components roughly fit to the correlation.  

\citet{ingalls07} observed \object{MBM-12} with OVRO interferometer and detected 1-5 milli-parsec scale clouds in the CO line wings.  They discussed that the velocity field of the cloud is made of an ensemble of tiny diluted turbulent cells.  

In the course of the natal cloud survey of the young stellar cluster \object{TW Hya association}, \citet{tachihara09} detected small clouds by the NANTEN telescope.  They have complex velocity structures and patchy morphology, barely resolved by the low spatial resolution observation ($2\farcm7$), but the line width is relatively narrow as $\sim 1$ km s$^{-1}$.  Some of them are, on the other hand, revealed to have small distance ($< 100$ pc), indicated by the optical Na absorption lines.  This ensures that the total molecular mass derived from CO within the surveyed area of 1.9 deg$^2$ is  $\la 3~M_\sun$.  
These results suggest that such small-scale cloud structures are common in the ISM.

\subsection{Nature of the small-scale structures and origin of interstellar turbulence}

As for the physical nature of the ubiquitous inter-stellar turbulence from diffuse to dense gas, the Kolmogorov model has been supposed to be a plausible mechanism.  According to the model,  turbulence is driven by a larger-scale wave such as a supernova explosion, stellar wind, and molecular outflows, and cascades into smaller-scale structures.  This cascade results in an energy spectrum  expressed by a power-law function as $E(k) \propto k^{-5/3}$ where $E$ is the energy of the turbulent flows and $k$ is the wave number.  Some observational results support this model \citep[e.g.,][]{armstrong95}, however, the driving mechanism is known to have a difficulty that the large-scale supersonic flows cause shock dissipation and most of the input energy is rapidly lost by radiation.  In order to solve this problem, many magnetohydrodynamic (MHD) simulations have been attempted with a help of the magnetic Alfv\'en wave.  Nevertheless, it is reported that the MHD turbulence decays within a timescale of the order of the flow crossing time even with strong magnetic fields \citep{maclow98,stone98}, because the cascade of the MHD turbulence happens within an eddy turnover time \citep{cho05}.

According to the two-phase medium model, on the other hand, the molecular clouds in turbulent interstellar medium are built up from small scale structures formed by thermal instability and their coalescence.  The ISM consist of two-phase gas, WNM and CNW, that can weaken the shock dissipation of the supersonic motion and prolong the decay time of turbulence \citep{inutsuka05}.  In the $\rho$ Ophiuchi cloud, CO and \ion{C}{1} gas is reported to be well mixed \citep{kamegai03}, which supports the idea of the two-phase medium.  If a cloud is interacting with an \ion{H}{2} region like the present case, the cloud surface is exposed to the ionization shock, that produces turbulence via the thermal instability.  The detection of the small scale structures with cold temperatures and high densities is strong support for the two-phase medium model.  

In addition to the results of density distribution of the original work by \citet{koyama02}, \citet{yamada07} further investigated the two-phase model including the radiative transfer calculation to estimate the temperature distribution as well as calculation of the line ratios of CO and \ion{C}{2} emissions.  They suggested that the small-scale CNM are formed from WNM via thermal instability having a small dense structure with minimum temperature of $\sim 50$ K, and the estimated $R_{3-2/1-0}$ is between 2 and 3, slightly higher than our result.  Their condition is, however, set to be lower density ($n{\rm (H_2)} \la 3000$ cm$^{-3}$), and for the case of higher density and lower temperature, $R_{3-2/1-0}$ is expected to be smaller.  

Note that the numerical simulations of \citet{koyama02} and \citet{yamada07} begin with uniform WNM compressed by passage of a shock wave.  
\citet{koyama02} showed that if the pre-shock gas is essentially one-phase diffuse WNM, the thermal instability induced by shock compression results in forming two-phase turbulent medium.  
However, the present case of our observed molecular cloud (CNM) is supposed to be clumpy (multi-phase) medium that preexists even before the UV ionization takes place.  This is because the UV source of \object{$\zeta$ Oph} is a run-away star, it has been irradiating the cloud only for a few $\times 10^5$ yrs, which is shorter than the timescale of thermal instability \citep{tachihara00b}.  In addition, the molecular CO gas seems to be well mixed with atomic \ion{C}{1} gas even in dense parts of the cloud \citep{kamegai03}, the ISM in this region is supposed to be clumpy two-phase medium even before the interaction with \object{$\zeta$ Oph}.  
On the other hand, using MHD simulations, \citet{inoue09b, inoue11} revealed, however, that passage of shock wave through the two-phase medium drives the ISM turbulence owing to vortex creation by shock-cloud interaction.  This mechanism does not require thermal instability, and hence, the turbulence is driven more effectively with shorter timescale than the gas cooling time, and even shorter than the crossing time of \object{$\zeta$ Oph}.  
It is also anticipated that under the strong UV radiation the ISM should have larger turbulence due to larger input energy causing larger line width \citep{tachihara00a, tachihara02, grischneder09}.  

In order to further study the origin of interstellar turbulence, high resolution interferometric observations in multi-transition CO lines at cloud boundaries are required.  Also for the investigation of physical properties of WNM, extensive \ion{C}{1} observation in the submm wavelength is highly desired.  Observational results brought by ALMA combined with ACA would enormously improve our understanding in the near future.  

\section{Summary}

In order to investigate the origin of interstellar turbulence and verify the theoretically proposed thermal instability model of two-phase medium, we have carried out CO $J=$1--0 and 3--2 observations toward a boundary between \object{LDN 204} dark cloud and an \ion{H}{2} region \object{Sh 2-27}.  The main conclusions are as follows.
\begin{enumerate}
\item By detailed investigation of spatial and velocity distribution of CO $J=$1--0, small-scale cloud structures have been discovered.  
\item They have characteristic morphologies such as clumpy, arc-like, and pillar-like structures, which appear in some velocity channels with typical size scale of a few $\times 1000$ AU to 12000 AU and the velocity dispersion of $\Delta V \sim 0.6$ km s$^{-1}$.  The systemic velocities of these 0.6 km s$^{-1}$ width component clouds have scatter of a few km s$^{-1}$.  
\item The spectral line profile taken in the middle of the main cloud, and the composite line profile of the entire cloud resemble single component broad-line spectra with velocity dispersion several times larger than the typical sound speed.  These facts suggest that the cloud is composed of tiny cloudlets with small internal velocity dispersion.  Because of their relative motion and overlapping effect on the line of sight, molecular clouds are observed to have supersonic turbulence as a whole.
\item Follow-up observations in CO $J=$3--2 for these small structures revealed that the intensity ratio of $R_{3-2/1-0}$ is close to unity for the 0.6 km s$^{-1}$ width component, which shows minimum kinetic temperature of $T_{\rm kin} > $ 25 - 100 K, relatively higher than typical dark clouds.  
\item From the LVG analysis, it is found that the 0.6 km s$^{-1}$ width components have average density of $\sim 10^{3-4}$ cm$^{-3}$, much higher than the entire cloud average density.  Owing to their small sizes, masses of the 0.6 km s$^{-1}$ width components are estimated to be $0.005 \la M \la 0.05~M_{\sun}$, well in the domain of brown-dwarf mass.
\item These results strongly support the theoretical thermal instability model of two-phase medium first proposed by \citet{koyama00, koyama02}.  We conclude that the interstellar turbulence is likely to be driven by thermal instability and the small-scale structures are the building blocks of the turbulent molecular clouds.

\end{enumerate}

\acknowledgements

We are grateful to staff members of the Nobeyama radio observatory and the ASTE 
support team.  
We thank Miyuki Murai for helping with the 45 m telescope observations and 
Takeshi Tsukagoshi for the ASTE observations.  
Nobeyama Radio Observatory is a branch of the National Astronomical Observatory of Japan, National Institutes of Natural Sciences.  
The ASTE project is driven by Nobeyama Radio Observatory (NRO), a branch of National Astronomical Observatory of Japan (NAOJ), in collaboration with University of Chile, and Japanese institutes including University of Tokyo, Nagoya University, Osaka Prefecture University, Ibaraki University, and Hokkaido University.  
Observations with ASTE were in part carried out remotely from Japan by using NTT's GEMnet2 and its partnet R\&E (Research and Education) networks, which are based on AccessNova collaboration of University of Chile, NTT Laboratories, and NAOJ.  
This work is financially supported by the Grants-in-Aid for the Scientific Research 
by the Ministry of Education, Science, Sports and Culture (No.~23540277 and 23740154).  
This publication is supported as a project of the Nordrhein-Westf\"alische
Akademie der Wissenschaften und der K\"unste in the framework of the
academy program by the Federal Republic of Germany and the state
Nordrhein-Westfalen.

%\newpage
\appendix
\section{Interstellar dust component of the cloud}

Besides the molecular and ionized gas components, dust is an important content of the ISM.  The emission, absorption and scattered light by the interstellar dust can provide morphological, geometrical and temperature information of the cloud.  

\begin{figure}
\epsscale{1}
\plotone{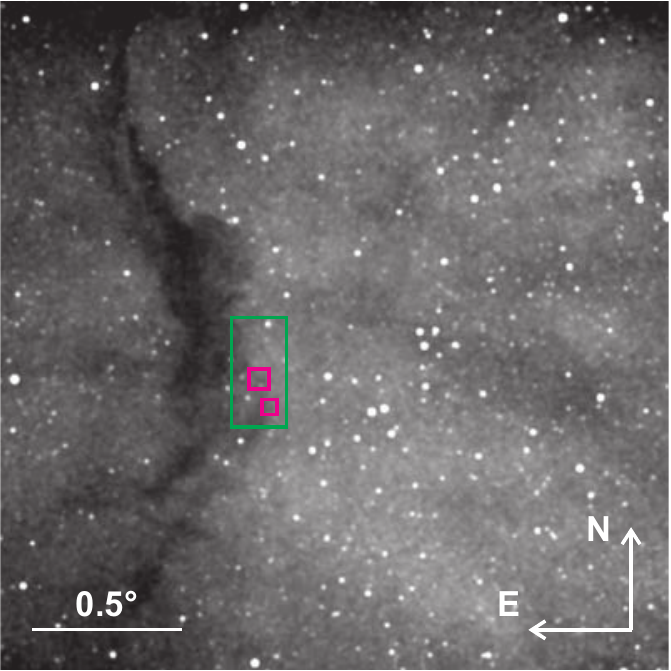}
\caption{H${\alpha}$ image ($2\fdg25 \times 2\fdg25$) of the LDN 204 cloud around the target region taken by the VYSOS-6 A telescopes at the Universit\"atssternwarte Bochum in Chile.  The regions observed by the 45m telescope and ASTE are designated by green and magenta rectangles, respectively.}
\label{Halpha}
\end{figure}

An H$\alpha$ image was obtained toward the entire cloud complex of \object{LDN 204} with VYSOS-6 A, a 150 mm telescopes of the Universit\"atssternwarte Bochum in Chile.  Ten images each with an integration time of 2 minutes were taken during March 24th and 25th 2012.  The images were taken with an H$\alpha$ filter whose central wavelength and the FWHM are 656 nm and 5 nm, respectively, and in dither-mode for bad-pixels removal.  After dark and bias subtraction and flatfielding the images were combined using a min/max-rejection. The reduced image was then binned ($8 \times 8$) and low-pass filtered.
In the observed region, the extinction of H$\alpha$ is not as prominent as in the main filament of \object{LDN 204}, but visible as a dark nebular near the edge of the spherical \ion{H}{2} region \object{Sh 2-27} (Fig.~\ref{Halpha}).  This ensures that the entire cloud is located in front of the Str{\" o}mgren sphere and gas is accelerated toward us by the UV light \citep{tachihara00b}.  

\begin{figure}
\epsscale{1}
\plotone{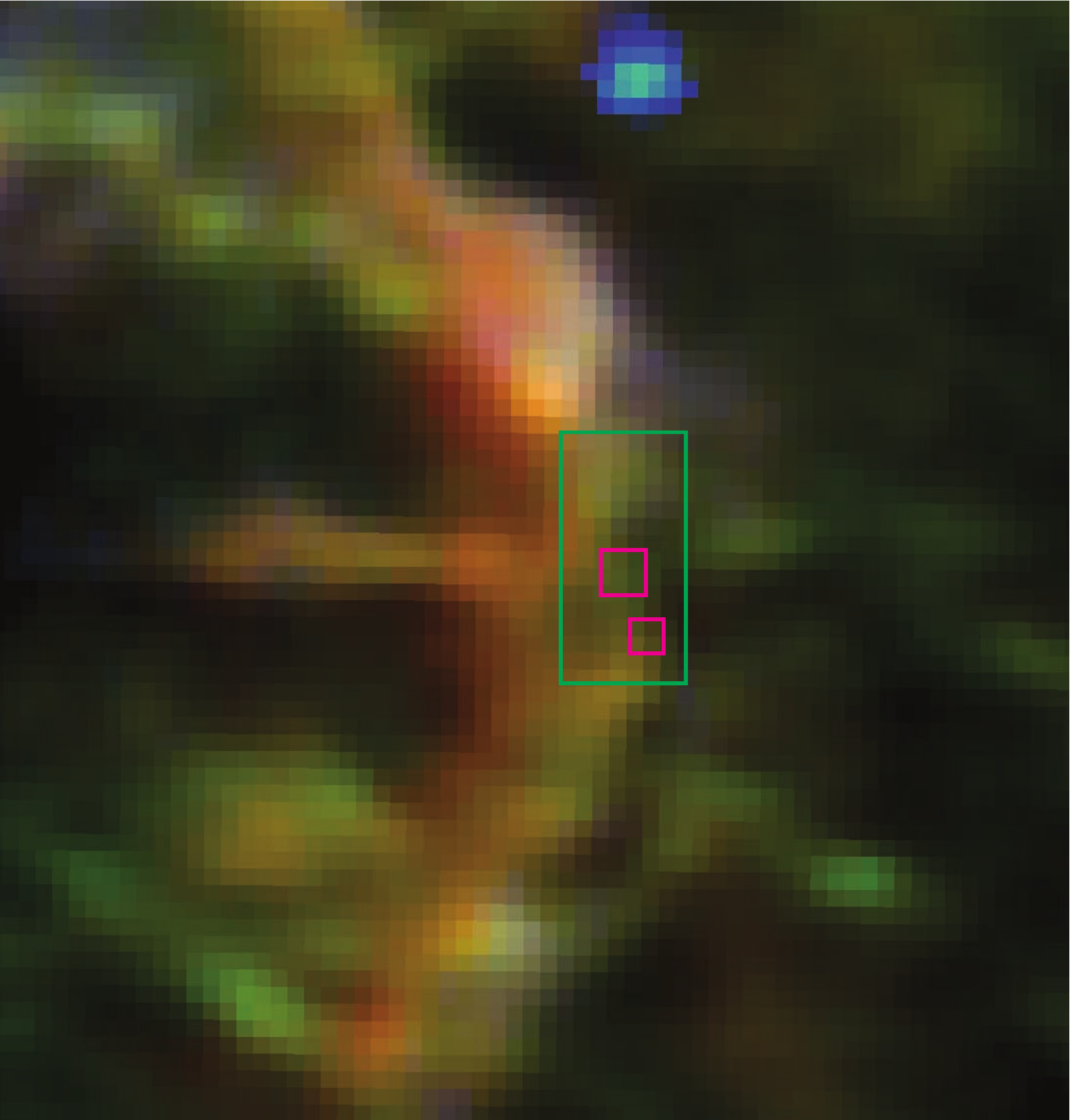}
\caption{IRIS 3-color composite image of the LDN 204 cloud, $100\arcmin \times 100\arcmin$ around the target region (12 $\mu$m in blue, 60 $\mu$m in green, and 100 $\mu$m in red).  The regions observed by the 45m telescope and ASTE are designated by green and magenta rectangles, respectively.}
\label{IRAS}
\end{figure}

The global distribution of the CO emission is compared with the dust thermal emission in 100 $\mu$m by \citet{tachihara00b}.  They are in good agreement in general, but the 100 $\mu$m emission is enhanced at the cloud surface irradiated by the UV light.  
The dust temperature distribution is represented by the color of far-IR thermal emission.  Fig.~\ref{IRAS} is the 3-color composite image of the entire region of \object{LDN 204} produced with the Improved Reprocessing of the IRAS Survey (IRIS) data\footnotemark.  The cloud surface is heated by UV radiation and the dust has relatively high temperature as demonstrated by the enhanced 60 $\mu$m emission over 100 $\mu$m.  The local high temperature causes the enhancement of 100 $\mu$m emissivity at the cloud surface relative to the CO line \citep{tachihara00b}.  The dense parts of the cloud are, on the contrary, in low-temperature indicated by relatively strong emission in 100 $\mu$m.  
\footnotetext{The IRIS data is obtained from http://irsa.ipac.caltech.edu/data/IRIS/}

\begin{figure}
\epsscale{1}
\plotone{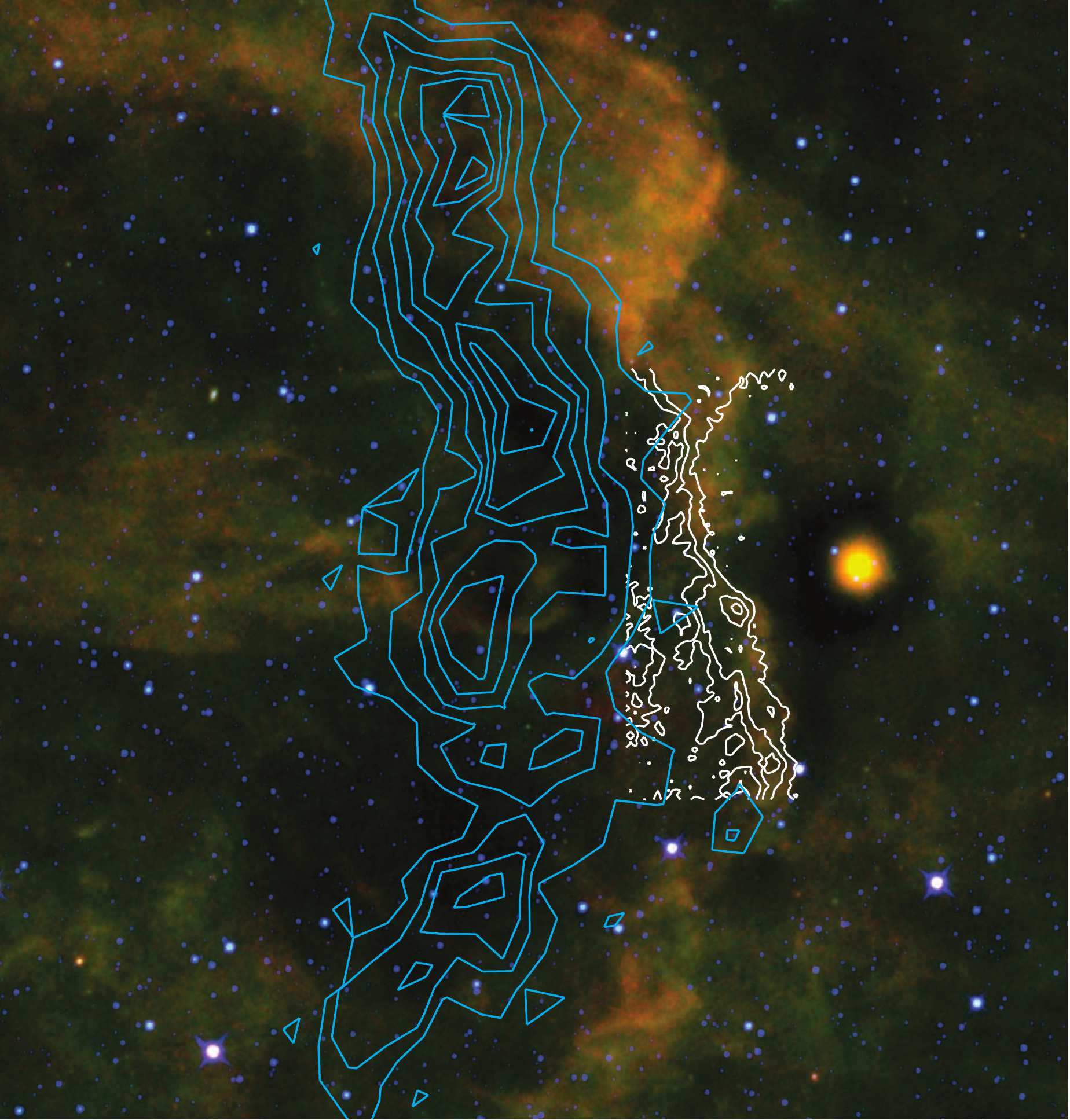}
\caption{Composite 3-color composite image of WISE (4.5 $\mu$m in blue, 12 $\mu$m in green, and 22 $\mu$m in red), 1 deg$^{2}$ around the LDN 204 cloud.  The bright yellowish object in the right is a ghost of V446 Oph, which is out of the image.  The white contours are the CO $J=$1--0 peak $T_{\rm MB}$ distribution by the present study with the 45m telescope, while those in cyan are the integrated intensity distribution of C$^{18}$O observed by \citet{tachihara00a} with the NANTEN telescope.}
\label{WISE}
\end{figure}

Recent preliminary release of the Wide-field Infrared Survey Explorer (WISE) data\footnotemark\ is compared with the CO distribution.  Fig.~\ref{WISE} shows 3-color composite image in the region around \object{LDN 204} using the band 4 (22 $\mu$m) in red, band 3 (12 $\mu$m) in green, and band 2 (4.5 $\mu$m) in blue.  The 22 $\mu$m emission is supposed to come from small dust particles not in thermal equilibrium, while the 12 $\mu$m one is rather dominated by the emission from polycyclic aromatic hydrocarbons (PAH) excited by UV light \citep[e.g.,][]{compiegne2011}.  These mid-IR emission in general delineate the cloud surface facing the \ion{H}{2} region, while the emission in 12 $\mu$m appears to be more spread than in 22 $\mu$m.  On the other hand, the dense part in the cloud traced by C$^{18}$O $J=$1--0 \citep{tachihara00a} appears to be a dark lane in mid-IR, unlike at 100 $\mu$m, slightly offset to the far side from \object{$\zeta$ Oph}.  The dust in the cloud is supposed to have high column density and low temperature in the dark lane.  The CO $J=$1--0 distribution of the present study traces the edge of the cloud where the column density is not so high but mid-IR is bright in emission.  
Besides the cloud's edge, diffuse mid-IR emission is extended outside the region detected in CO.  It suggests that considerable amount of molecular gas is photo-evaporated by the UV radiation but the warm dust grains remain at the cloud surface.  This feature is, on the contrary, not clearly visible in the absorption feature of H$\alpha$, implying that the warm dust cloud has relatively low column density.  
\footnotetext{The WISE data is obtained from http://irsa.ipac.caltech.edu/Missions/wise.html}

%\newpage

\end{document}